# Fiber-integrated microcavities for efficient generation of coherent acoustic phonons

O. Ortiz[1*], F. Pastier[2*], A. Rodriguez[1], Priya[1], A. Lemaitre[1], C. Gomez-Carbonell[1], I. Sagnes[1], A. Harouri[1], P. Senellart[1], V. Giesz[2], M. Esmann[1~], N.D. Lanzillotti-Kimura[1+]

[1] Université Paris-Saclay, CNRS, Centre de Nanosciences et de Nanotechnologies (C2N), 10 Boulevard Thomas Gobert, 91120 Palaiseau, France

[2] Quandela SAS, 10 Boulevard Thomas Gobert, 91120 Palaiseau, France

* these authors contributed equally

~ martin.esmann@c2n.upsaclay.fr
+ daniel.kimura@c2n.upsaclay.fr

**Abstract:** *Coherent phonon generation by optical pump-probe experiments has enabled the study of acoustic properties at the nanoscale in planar heterostructures, plasmonic resonators, micropillars and nanowires. Focalizing both pump and probe on the same spot of the sample is a critical part of pump-probe experiments. This is particularly relevant in the case of small objects. The main practical challenges for the actual implementation of this technique are: stability of the spatio-temporal overlap, reproducibility of the focalization and optical mode matching conditions. In this work, we solve these three challenges for the case of planar and micropillar optophononic cavities. We integrate the studied samples to single mode fibers lifting the need for focusing optics to excite and detect coherent acoustic phonons. The resulting excellent reflectivity contrast of at least 66% achieved in our samples allows us to observe stable coherent phonon signals over at least a full day and signals at extremely low excitation powers of 1µW. The monolithic sample structure is transportable and could provide a means to perform reproducible plug-and-play experiments.*

Coherent phonon generation by optical pump-probe experiments[1–3] has enabled the study of acoustic properties at the nanoscale in planar heterostructures[4–12], plasmonic resonators[13–17], cells[18], micropillars[6,19–21] and nanowires[22].

All these experiments rely on the optical mode matching between the incident pump and probe laser fields and the optical modes of the structure under study. The efficient generation of coherent acoustic phonons relies on an efficient coupling of the pump field into the system, while the sensitive detection of phonons requires an efficient coupling of the probe to the optical mode undergoing a phonon-induced modulation. The main practical challenges for the actual implementation of this technique are therefore: 1) stability 2) reproducibility and 3) high power densities limiting the range of compatible samples[23].

The implementation of these experiments, also known as time domain Brillouin scattering[24], usually requires a long mechanical delay line. In terms of stability, this implies issues due to changes in beam pointing, spot size and laser power over the scan of the delay line. The diversity of possible experimental configurations makes identical measurements in different laboratories virtually impossible. Even more, successive realignments of a single setup might result in slightly different excitation and collection mode matching. These shortcomings have so far been a roadblock in establishing pump-probe as a quantitative spectroscopy tool for nanoacoustics.

A partial solution mainly addressing stability considerations is provided by the asynchronous optical sampling (ASOPS) technique by removing the need for a mechanical delay line [23]. However, alignment and power requirements are still inherent of the experiment. In this work, we integrate fibered systems to eliminate any optical alignment on the sample, thus simultaneously solving the three aforementioned main problems. We glued single mode fibers to a planar optophononic microcavity and to an optophononic micropillar [6,19–21,25–28]. The integration with fibers might establish the missing link between high frequency acoustic phonon engineering [29–33] and stimulated Brillouin scattering in structured optical fibers [34–38].

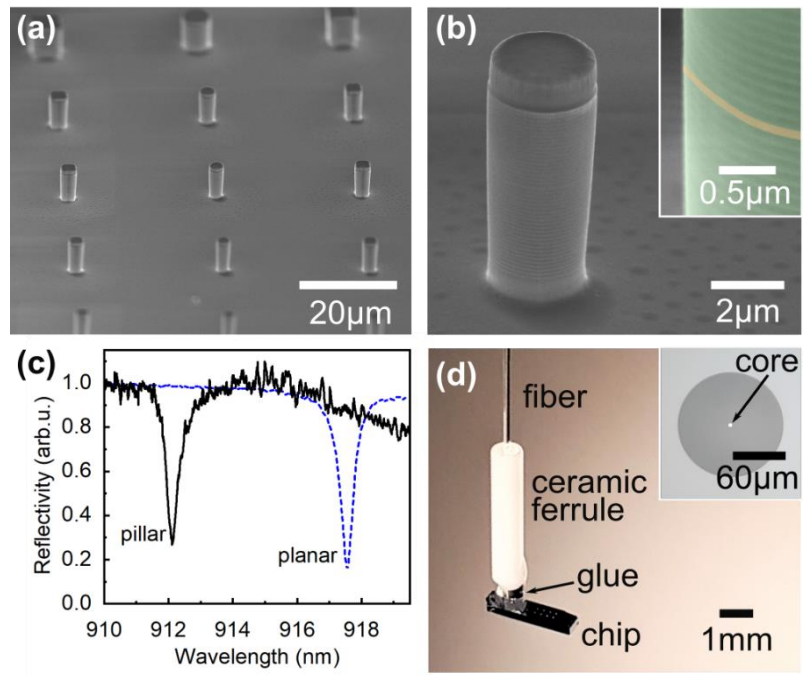

*Figure 1 Integrated fibered system for pump probe experiments. (a) SEM image of a field of GaAs/AlAs micropillars presenting different shapes and sizes. (b) SEM image of an individual micropillar cavity of 2.8 μm diameter. The inset shows a detail of the DBRs (green) and the central GaAs spacer (orange). (c) Optical reflectivity spectrum of the fibered micropillar after we permanently glued it to the core of a single mode fiber (solid black). The fundamental optical mode appears as a dip at 912.1 nm. The spectrum of the connected planar cavity (dashed blue) shows a fundamental optical mode at 917.6 nm (d) Final device with one individual micropillar on the chip coupled to the single mode fiber. A ceramic ferrule provides additional mechanical stability. Inset: microscope image of fiber core and cladding.*

Figure 1a shows a field of micropillars obtained by electron beam lithography and inductively coupled plasma etching of a planar microcavity. The planar cavity is formed by two (λ/4 $Ga_{90}Al_{10}As$, λ/4$Ga_5Al_{95}As$) distributed Bragg reflectors (DBRs) enclosing a λ/2 spacer of GaAs. 19 (23) DBR layer pairs on the air (GaAs substrate) side result in a balanced optical cavity. We grew the sample by molecular beam epitaxy with a thickness gradient across the wafer, which implies a spatially dependent optical resonance. Figure 1b presents a single circular micropillar of 2.8 μm diameter. The inset shows a close-up of the central pillar section. The DBR structures and the cavity spacer are clearly visible. This micropillar presents an optical mode at 912.1 nm evidenced as a reflectivity dip (black line in Fig. 1c) whereas the planar microcavity has an optical resonance at 917.6 nm (dashed blue line in Fig. 1c). This wavelength difference is due to the spatial gradient across the wafer. For the individual micropillar, we spatially overlap the core of a single mode optical fiber (Thorlabs 780HP, core diameter 4.4 μm) with its top surface [39–42]. To this end, we monitor the micropillar reflectivity dip through the fiber while adjusting the sample chip position on a 3D piezo motor with submicron precision. In contrast, for the planar cavity a precise lateral alignment of the fiber core is not necessary. Once spatial alignment is completed, we permanently glue the fiber to the sample chip (UV-curing glue Norland

NOA 81). During the curing process of the glue the optical mode frequency undergoes only minimal shifts (<0.1%). Figure 1d presents a picture of a finished device. A ceramic ferrule (white tube) provides additional mechanical stability. The mode overlap of the fiber to the microresonators achieved during the gluing is permanent and stable. No further alignment steps are required.

Our measurement setup for the integrated samples is based on a fiber beam splitter (see schematics in Fig. 2a). To evaluate the loss budget in the system, we identify three main contributions: The fiber beam splitter itself, the fiber coupler (FC), and the fiber-to-sample coupling. We define the coupling efficiency as the ratio of collected power in arm 3 of the beam splitter to the input power injected from arm 1. Based on standard values for the beam splitter transmission (50%) and FC transmission (80%), we obtain a theoretical maximum coupling efficiency of 16% for the planar cavity at a wavelength outside the cavity mode. Note that this coupling efficiency takes into account both input and output coupling. The measured coupling efficiency is 12.7% with the main potential source of additional losses being the fiber-to-sample coupling (distance and angle). In the case of the micropillar an additional loss is the geometrical overlap between the fiber core and the micropillar top surface (40%). We obtain a theoretical maximum coupling efficiency of 2.5% for a wavelength outside the cavity mode. We measure a coupling efficiency of 0.8%, again mainly determined by additional potential misalignment of distance and angle between pillar and fiber core.

The reflectivity contrast for the optical mode is the additional parameter which determines the sensitivity to detect acoustic phonons, since the detected magnitude is usually the transient reflectivity change $\Delta R/R$. In the present cases, the reflectivity contrast is 66% for the micropillar and 80% for the planar cavity.

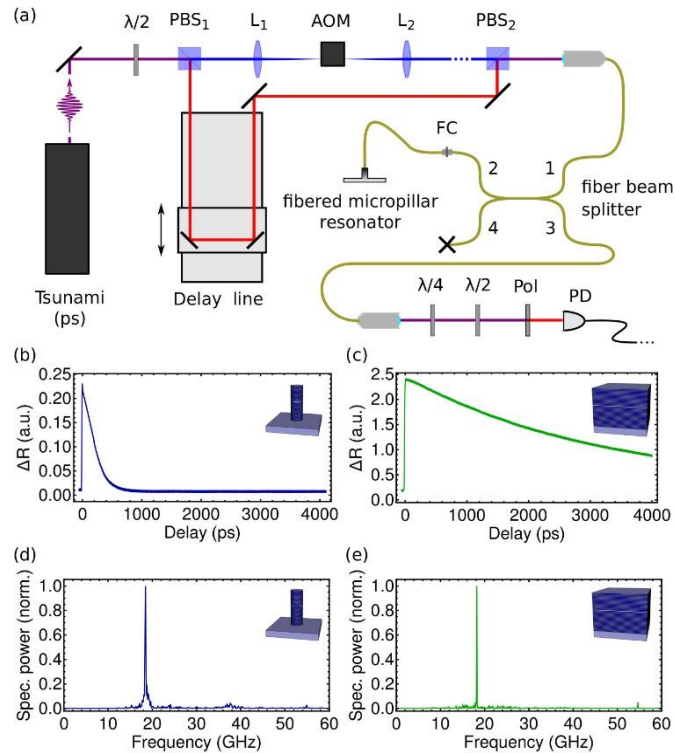

*Figure 2 : (a) Schematic of the pump-probe setup used to characterize the optophononic response of fiber-coupled microcavities. We replaced the standard focalization optics by a fibered system. (b,c) Differential reflectivity time traces of the micropillar and planar resonators. (d,e) Fast Fourier transform (FFT) of the experimental time traces featuring the acoustic resonance at 18.43 GHz (18.22 GHz) for the micropillar (planar resonator) and its third harmonic. Powers were 500μW for the pump and 700μW for the probe.*

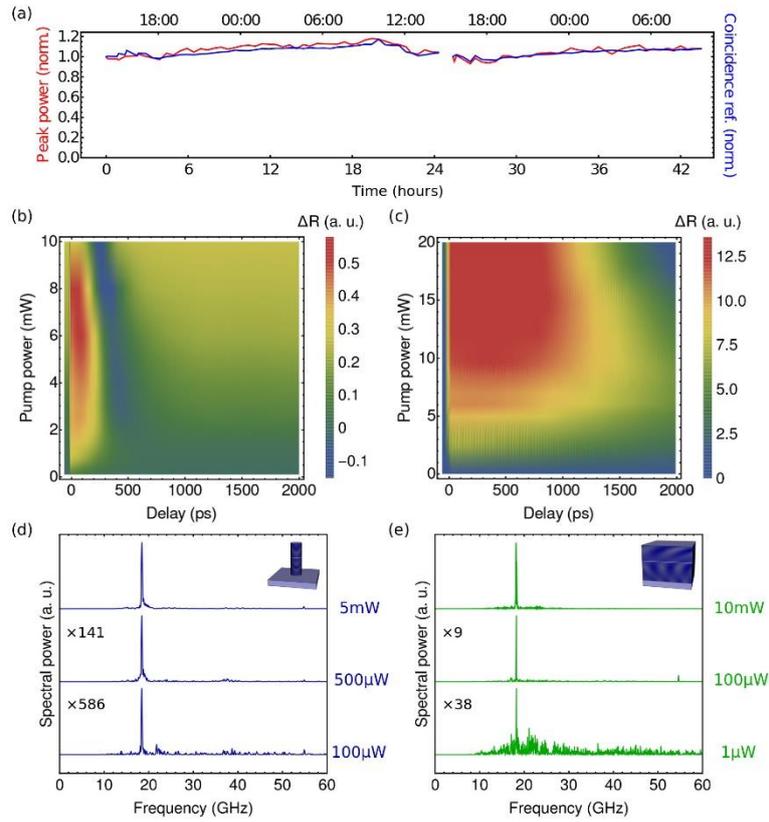

*Figure 3 : (a) Stability analysis of the fibered micropillar resonator over 42h; amplitude of the time trace at zero delay (blue) and spectral peak power of the acoustic cavity mode at 18.43 GHz (red). (b) Pump-probe traces as a function of pump power for the fibered micropillar cavity and (c) for the planar resonator. The results shown in the color maps are the interpolation of 19 and 25 individual measurements for the micropillar and the planar cavity, respectively. We kept the incident probe power constant at 700µW. (d) and (e) present corresponding Fourier spectra at selected low and high pump powers for the micropillar and planar cavity, respectively. Factors on the left of each plot indicate vertical magnification with respect to the top curve. Note that the lowest pump power used was 1µW for the planar cavity.*

To investigate the fibered devices, we have implemented the degenerated pump-probe setup sketched in Figure 2a. A Ti:Sapphire laser (Spectra Physics Tsunami) delivers pulses of 3.8 ps full width at half maximum (FWHM) with 80 MHz repetition rate and at a central wavelength of 912.08 nm for the micropillar (917.58 nm for the planar resonator). We divide the beam into cross-polarized pump and probe beams. A mechanical delay line (3.6 m maximum delay) delays the probe with respect to the pump. We use an acousto-optical crystal to modulate the pump (800 kHz modulation frequency, 50% duty cycle) enabling synchronous detection of pump-induced changes in the probe signal. Both beams recombine and enter arm 1 of a fiber beam splitter, while arm 2 connects to the sample. The same arm 2 collects reflected pump and probe signals and sends them to a photodetector through arm 3. A power meter at arm 4 of the beam splitter monitors input powers. Since pump and probe are equal in wavelength and are contained in the same spatial mode of the fiber, the synchronous detection critically relies on polarization filtering of the pump probe signals. We compensate any rotation of polarization in the fibers by means of a combination of a quarter- and half-wave plate. A linear polarizer then filters out the pump before the signal reaches the photodetector. To minimize random fluctuations of polarization rotation within the fibers, we mechanically stabilize all fibers in the system.

Figures 2b and c display typical pump-probe differential reflectivity time traces measured in fiber-coupled planar and micropillar resonators. Pump powers of 500 µW and probe powers of 700 µW were measured in arm 4 of the fiber beam splitter. At zero delay, we observe a sharp pump-induced reflectivity change followed by a slow decay of the signal. While this decay happens on a time scale of 4ns for the planar resonator the decay only takes approx. 0.5 ns for the micropillar. This difference might be attributed to faster recombination of electronic excitations at surface defects[43,44]. The coherent phonon-induced oscillations superimposed with this electronic response are not evident to the naked eye. Only by Fourier transforming the time traces, we obtain the coherent acoustic phonon spectra in Figure 2d and e. We observe a main peak around 18.43 GHz (18.22 GHz) corresponding to the fundamental acoustic cavity mode of the micropillar (planar resonator) and additional peaks at 58 GHz corresponding to their third harmonic. The acquisition time for each of these measurements was just five minutes for the micropillar as well as for the planar cavity. In free space, similar experiments using a mechanical delay line usually take on the order of one hour and higher laser powers.

Our fiber-coupled samples provide inherently stable mode overlap with no further alignment involved. The measurements thus only depend on the characteristics of the laser and the stability of the external fiber coupling. To assess this stability, we performed long-term pump-probe measurements over the course of 42h only adjusting the incident power (measured in arm 4 of the fiber beam splitter) once after 24h to compensate for drift of the laser. Figure 3a presents the amplitude of the differential reflectivity at zero delay (blue) and the spectral peak power of the mode at 18.43 GHz (red) over time. Both are normalized to the first datapoint. The curves show variations of 15% dominated by a 24h temperature cycle in the laboratory. As long as the coupled laser characteristics remain constant, the resulting pump-probe traces are the same. This means, that by transferring the integrated sample exactly the same pump-probe measurement could be performed in any other experimental pump-probe setup.

Figure 3b and c show intensity maps of pump-probe traces as a function of pump power for micropillar (b) and planar (c) resonators. In these experiments, we kept the incident probe power constant at 700µW. Figures 3d and e present corresponding Fourier spectra at selected low and high pump powers. Note that for the case of the planar cavity we observe phonon signals down to 1µW of pump power (100 µW for the pillar). The essential features of the pump-probe spectra remain unchanged with pump power. We measured these powers in arm 4 of the fiber beam splitter, equivalent to the power injected into the FC coupler at arm 2. Similar phonon dynamics have been reported before, yet for much higher pump powers in the range of 5mW [20].

The observation of complex optical mode dynamics is mainly related to the efficient pump injection whereas the observation of coherent phonons at extremely low pump powers indicates an efficient probe coupling. This implies that our mode matching for both pump and probe enables a new generation of pump-probe experiments where low powers are required.

The impulsive generation of coherent acoustic phonons has a broad of applications spanning from non-destructive testing to the study of quantum optomechanical systems. Experimental challenges typically include mechanical stability and reproducing equivalent experimental conditions over time and between setups. Furthermore, high power densities are typically needed limiting the range of accessible samples. We proposed a solution to these three challenges consisting of integrating the studied sample in a fibered system. We showed that gluing optophononic semiconductor cavities to a single mode fiber lifts the need for focusing optics to excite and detect coherent acoustic phonons. The mode matching achieved in our samples allowed us to observe stable signals over at least a full day and coherent phonon signals at extremely low excitation powers of 1μW. The monolithic sample structure is transportable making it a potential means to perform reproducible plug-and-play pump-probe experiments in individual microstructures.

Our results could enable a new generation of nanophononic experiments in systems such as planar heterostructures (quantum wells, multilayers, lasers, quantum dots, magnetic thin films[7], piezo-electric materials, 2D materials), organic materials (not resistant to high powers usually needed in coherent phonon generation experiments) single cells and metallic nanostructures (antennas, metamaterials, metasurfaces). In our case, we distinguished pump and probe signals based on polarization. However, the proposed technique would also allow two-color experiments as a straightforward extension.

**Acknowledgements:**

The authors thank P. Voisin for enlightening discussions. The authors acknowledge funding by the European Research Council Starting Grant No. 715939, Nanophennec. This work was supported by the European Commission in the form of the H2020 FET Proactive project TOCHA (No. 824140). The authors acknowledge funding by the French RENATECH network and through a public grant overseen by the ANR as part of the "Investissements d'Avenir" program (Labex NanoSaclay Grant No. ANR-10-LABX-0035). M.E. acknowledges funding by the Deutsche Forschungsgemeinschaft (DFG, German Research Foundation) Project 401390650. F. P. and V. G. acknowledge support from the IAD - ANR program (Project ANR-18-ASTR-0024 LIGHT).